\documentclass[reprint,
superscriptaddress,
amsmath,amssymb,
aps,
longbibliography,
prl,
hidelinks]{revtex4-1}

\usepackage{graphicx}
\usepackage{soul}
\usepackage{dcolumn}
\usepackage{bm}
\usepackage{hyperref}
\usepackage{xr-hyper}
\usepackage{xcolor}
\usepackage[normalem]{ulem}

\usepackage{mathtools}
\newcommand\underrel[3][]{\mathrel{\mathop{#3}\limits_{%
			\ifx c#1\relax\mathclap{#2}\else#2\fi}}}

\begin{document}

\preprint{APS/123-QED}

\title{Active Bound States Arise From Transiently Nonreciprocal Pair Interactions}
	\author{Luca Cocconi}
	\thanks{These authors contributed equally.}
	\affiliation{Department of Mathematics, Imperial College London, South Kensington, London SW7 2AZ, United Kingdom}
	\affiliation{The Francis Crick Institute, London NW1 1AT, United Kingdom}
	\affiliation{Max Planck Institute for Dynamics and Self-Organization (MPIDS), 37077 G\"{o}ttingen, Germany}
		
	\author{Henry Alston}
	\thanks{These authors contributed equally.}
	\affiliation{Department of Mathematics, Imperial College London, South Kensington, London SW7 2AZ, United Kingdom}
	
	\author{Thibault Bertrand}%
	\email{t.bertrand@imperial.ac.uk}
	\affiliation{Department of Mathematics, Imperial College London, South Kensington, London SW7 2AZ, United Kingdom}

\date{\today}

\begin{abstract}
\noindent Static nonreciprocal forces between particles generically drive persistent motion reminiscent of self-propulsion. Here, we demonstrate that reciprocity-breaking fluctuations about a reciprocal mean coupling strength are sufficient to generate this behavior in a minimal two-particle model, with the velocity of the ensuing \textit{active} bound state being modulated in time according to the nature of these fluctuations. To characterize the ensuing nonequilibrium dynamics, we derive exact results for the time-dependent center of mass mean-squared displacement and average rate of entropy production for two simple examples of discrete- and continuous-state fluctuations. We find that the resulting dimer can exhibit unbiased persistent motion akin to that of an active particle, leading to a significantly enhanced effective diffusivity.

\end{abstract}

\maketitle


Newton's third law states that microscopic forces respect action-reaction symmetry, yet many examples of nonreciprocal effective interactions have been identified in living and reactive systems. These range from classical predator-prey \cite{Mobilia2007, Biktashev2009} and activator-inhibitor \cite{Theveneau2013} models to interactions mediated by a nonequilibrium medium \cite{Hayashi2006, Saha2019, Soto2014, AgudoCanalejo2019}. Nonreciprocity also arises in systems with asymmetric information flows \cite{Durve2018} and memory effects \cite{Loos2020,Loos2019}. 

The breaking of reciprocal symmetry in many-body systems generates fundamentally nonequilibrium dynamics at the collective scale \cite{Ivlev2015, Saha2020, You2020, Fruchart2021,Zhang2022a}. Most strikingly, an imbalance in effective physical forces between particles can drive persistent motion, reminiscent of self-propulsion \cite{You2020}. Motile particle clusters \cite{Soto2014,Niu2018,Saha2019,AgudoCanalejo2019,Lavergne2019} and self-propelling droplets \cite{Meredith2020} have been experimentally realized in systems with constant nonreciprocal couplings. Furthermore, the thermodynamic implications of reciprocity-breaking were studied in several theoretical models \cite{Loos2020,You2020}.

In principle, the introduction of temporal fluctuations in nonreciprocal interactions would provide a mechanism to control the propulsion speed and direction of the ensuing dynamical phase. Such fluctuations may for instance arise generically in physical systems through dynamic properties in the nonequilibrium medium that mediates interactions. Examples include the concentration of so-called \textit{doping agents} in chemically interacting particle systems \cite{AgudoCanalejo2019} or of a surfactant in an experimental set-up of self-propelled liquid droplets which allowed for the reversal of the direction of motion \cite{Meredith2020}. In a recent study, active motion was shown to emerge from the application of an external random magnetic field on nanoparticle dimers \cite{LuisHita2022}.

Fluctuating {\it reciprocal} interactions in many-body systems lead to nonequilibrium, dissipative structures \cite{Alston2022a} and dynamics \cite{Bonazzi2018}. We have previously studied the thermodynamic implications of these interactions in \cite{Alston2022b}, where we obtained analytically the non-zero average rate of entropy production in a variety of minimal setups. Though static nonreciprocal couplings have been studied in a similar manner \cite{Loos2020, Zhang2022a}, a complete thermodynamically-consistent picture for dynamic, nonreciprocal interactions is key to the analysis of important reactive, active and living processes.

In this Letter, we consider a minimal two-particle model of fluctuating nonreciprocal forces. We fix the interactions to be reciprocal on average, yet we let them break the action-reaction principle transiently through temporal fluctuations in the interaction strengths, isolating the impact of reciprocal-symmetry-breaking fluctuations on the collective dynamics and thermodynamic properties of the system. We show that these systems can exhibit collective motion reminiscent of active particles, thus we refer to the resulting 2-particle dimers as \textit{active bound states}. For particular choices of the fluctuations and in the presence of steric repulsion, the ensuing dynamics can be mapped onto those of \textit{Run-and-Tumble} \cite{Zhang2022b,GarciaMillan2021, Solon2015} and \textit{Active Ornstein-Uhlenbeck} \cite{Martin2021, Bothe2021} particles. 

\textit{Minimal 2-particle model. ---} We consider a pair of Brownian particles in the overdamped limit with positions $x_1(t), x_2(t) \in \mathbb{R}$ and diffusivity $D_x$. Each particle is confined in a harmonic potential with time-dependent stiffness $k_{1,2}(t)$ generated by the other particle. The governing equations then take the form
\begin{subequations}
\begin{align}
    \dot{x}_1 &= -k_1(x_1 - x_2) - \partial_{x_1}U_{r}\big(|x_1-x_2|\big) + \sqrt{2 D_x}\xi_1 \\
    \dot{x}_2 &= -k_2(x_2 - x_1) - \partial_{x_2}U_{r}\big(|x_1-x_2|\big) + \sqrt{2 D_x}\xi_2
\end{align}
\label{eq:2p_sys}%
\end{subequations}
where $\xi_{1,2}(t)$ are uncorrelated zero-mean, unit-variance Gaussian white noises and $U_{r}(r)$ is a short-range, reciprocal, purely repulsive potential. 
For the time being, we focus on the case $U_{r} \equiv 0$, for which a number of closed form exact results can be derived. We will later show that introducing steric interactions strongly enhances the observed nonequilibrium behavior. Here, we consider binding potential stiffnesses of the form $k_i(t) = \bar{k} + \kappa_i(t)$, where $\bar{k} > 0$ is a constant mean stiffness introduced to ensure that the two particles remain in proximity of each other and $\kappa_{1,2}(t)$ are governed by zero-mean Markov processes setting the stiffness fluctuations \cite{Yuan2003,Alston2022b}. 

Through a change of variables to the center of mass $x=(x_1+x_2)/2$ and interparticle displacement $y=x_1-x_2$ coordinates, we can re-write Eq.\,(\ref{eq:2p_sys}) as
\begin{subequations}
\begin{align} 
    \dot{x}(t) &= -\frac{1}{2} \psi(t) y(t) + \sqrt{D_x} \xi_x(t) \label{eq:eom_x}\\
    \dot{y}(t) &= - (\varphi(t) + 2 \bar{k}) y(t) + \sqrt{4 D_x} \xi_{y}(t) \label{eq:eom_y}
\end{align}
\label{eq:2p_sys_ref}%
\end{subequations}
where $\xi_{x, y}(t)$ are again uncorrelated zero-mean unit-variance Gaussian white noise terms (see details in \footnote{Supplemental material}). In writing Eq.\,(\ref{eq:2p_sys_ref}), we have also defined the \textit{stiffness asymmetry} $\psi(t)=\kappa_1(t)-\kappa_2(t)$ and the \textit{total stiffness fluctuations} $\varphi(t)=\kappa_1(t) + \kappa_2(t)$. Note that $\psi(t)\ne 0$ is the signature of {\it broken reciprocal symmetry}. 

We study here both the dynamics and thermodynamics of these two-particle bound states. To quantify their collective dynamics, we derive exact analytical expressions for the time-dependent mean-squared displacement (MSD) of their center of mass, $\left\langle (x(t)-x(0))^2\right\rangle$ (hereafter, setting $x(0)=0$ by translational symmetry). From Eq.\,(\ref{eq:eom_x}), this MSD can be expressed in terms of the correlator $\langle \psi(s)\psi(s')y(s)y(s')\rangle$; {as shown in \cite{Note1}}, the decoupling between the dynamics of $y$ and $\psi$ allows us to factorize it and we write
\begin{equation}\label{eq:MSD_general}
   \left \langle x^2(t)\right\rangle = D_xt+ \frac{1}{4}\int_0^t ds\,\int_0^t ds' \, \langle \psi(s) \psi(s') \rangle \langle y(s) y(s') \rangle.
\end{equation}

The presence of nonreciprocal and fluctuating interactions drives our two-particle bound states out of equilibrium; to quantify this nonequilibrium behavior, we compute the entropy production rate at the level of Eq.\,(\ref{eq:2p_sys_ref}). We generically expect three contributions, respectively stemming from (i) the dynamics of the center of mass, (ii) the dynamics of the interparticle displacement and (iii) the stochastic dynamics of the stiffness fluctuations (see \cite{Note1} for a detailed derivation). 

Firstly, the center of mass moves following a drift-diffusion process with a time-dependent drift $v(t) = -\psi(t)y(t)/2$ {and diffusivity $D_x/2$}; this contribution to the entropy production rate thus takes the form $\lim_{t\rightarrow\infty}\dot{S}_i^{(x)} =2\langle v^2(t)\rangle/D_x$ \cite{Seifert2012}. Secondly, the dynamics of the interparticle displacement $y(t)$ can be mapped onto those of a single Brownian particle subject to diffusion in a fluctuating harmonic potential, $U(y, t) = (\varphi(t)+2\bar{k})y^2/2$, a case which we previously studied in \cite{Alston2022b}. Finally, a third contribution may come from the two-dimensional Markov process $(\varphi,\psi)$ governing the stiffness dynamics, should it not satisfy detailed-balance. Here, we only consider stiffness fluctuations generated by equilibrium processes and this last contribution thus vanishes. The total rate of entropy production can then be written as \cite{Cocconi2020,Alston2022b}
\begin{equation}\label{eq:EP_general}
    \lim_{t\rightarrow\infty}\dot{S}_i = \lim_{t\rightarrow\infty}\bigg(\dot{S}_i^{(x)} + \dot{S}_i^{(y)} \bigg)= \frac{\langle \psi^2y^2\rangle}{2D_x} + \lim_{t\rightarrow\infty}\dot{S}_i^{(y)}.
\end{equation}
In what follows, we consider two examples of specific prescriptions for the governing stochastic dynamics of the stiffness fluctuations $\kappa_{1,2}(t)$ and show that transiently nonreciprocal pair interactions lead to persistent motion of the center of mass $x(t)$, akin to that of an active particle.

\textit{Continuous fluctuations in interaction potentials. --- } Suppose that the two stiffness fluctuations follow correlated zero-mean Ornstein-Uhlenbeck processes with rate $\mu$ and diffusivity $D_\kappa$, 
\begin{equation}
	\dot{\kappa}_{i} = -\mu \kappa_{i}  + \sqrt{2D_\kappa}\bar{\eta}_{i}(t), \quad i \in \{1,2\} \label{eq:eom_AOU}
\end{equation}
where $\bar{\eta}_{1,2}(t)$ are zero-mean white noises satisfying
\begin{equation}
    \langle \bar{\eta}_{i}(t)\bar{\eta}_{j}(t')\rangle = C_{ij} \delta(t-t'),\quad\mathrm{with} \quad {\bf C} = \begin{pmatrix} 1 & \theta \\ \theta & 1 \end{pmatrix}
\end{equation}
where ${\bf C}$ is the symmetric covariance matrix and $\theta \in[-1, 1]$ quantifies how correlated the stiffness fluctuations are.

The governing equations for the stiffness asymmetry $\psi(t)$ and total stiffness fluctuations $\varphi(t)$ then take the form 
\begin{subequations}\label{eq:eom_cov}\begin{align}
	\dot{\psi}(t) &= - \mu \psi(t) + \sqrt{4D_\kappa(1-\theta)} \eta_\psi(t) \\
	\dot{\varphi}(t) &= -\mu\varphi(t) + \sqrt{4D_\kappa(1+\theta)} \eta_\varphi(t)\label{eq:varphi_dynamics_cont}
\end{align}\end{subequations}
where $\eta_{\psi,\varphi}(t)$ are now \textit{uncorrelated}, zero-mean unit-variance Gaussian white noise terms \cite{Note1}. In each of the two limits $\theta=\pm 1$, one of the noise terms disappears. For all $\theta>-1$, the interparticle displacement behaves as a Brownian particle in a confining potential with a stiffness that itself follows an Ornstein-Uhlenbeck process with mean $2\bar{k}$ and variance $2D_\kappa(1+\theta)/\mu$.

\begin{figure}
    \centering
    \includegraphics[scale=1]{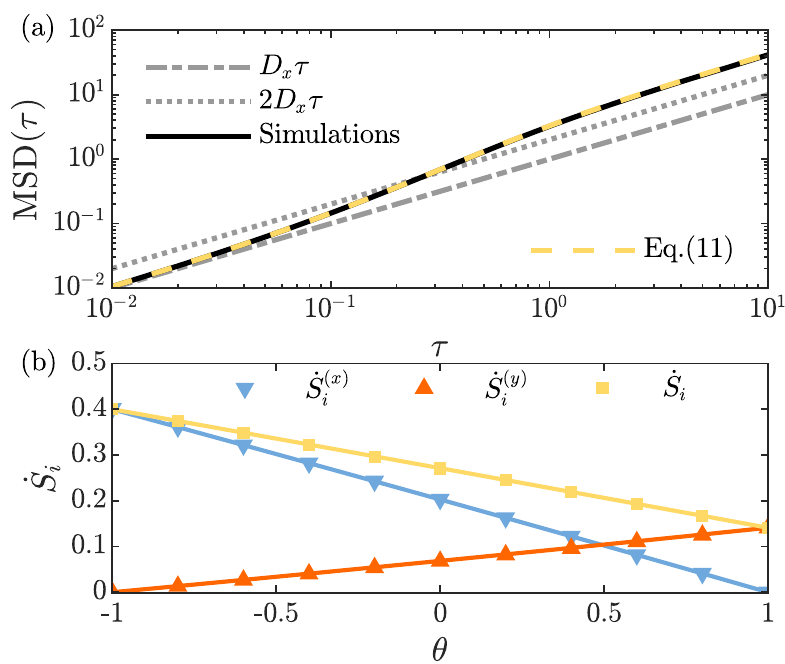}
    \caption{\textit{MSD and entropy production rate for correlated continuous fluctuations  ---} (a) MSD for the active bound states as given by Eq.\,(\ref{eq:fullmsd}), where we observe transient ballistic scaling implying persistent motion and an effective diffusion coefficient larger than that of an isolated particle. Here $D_\kappa=5$ and $D_x=\bar k = \mu = 1.$ (b) Rate of entropy production for the active bound state, made up of two contributions as identified in Eq.\,(\ref{eq:fullEP}), for $\bar k=5$,  $D_x=1$ and $D_\kappa = \mu = 10.$ Symbols are from numerical simulations (see \cite{Note1} for details) and solid lines are evaluated using Eqs.\,(\ref{eq:fullEP}) and (\ref{eq:fullvar}).}
    \label{fig:AOUP_EP}
\end{figure}

The dynamics for $\varphi$ and $\psi$ are independent, which implies that $\langle \psi^2 y^2\rangle = \langle \psi^2 \rangle \langle y^2 \rangle$ factorises in the second term of Eq.\,\eqref{eq:EP_general}.
Both contributions to the entropy production rate thus can be written in terms of the variance of the interparticle displacement, assuming that the latter is finite; the first is obtained by the results of \cite{Alston2022b}, while the second is deduced from the knowledge of the correlator for $\psi$:
\begin{equation}\begin{aligned}\label{eq:fullEP}
        \lim_{t \to \infty} \dot{S}^{({\rm cont.})}_i(t) &= \frac{\bar{k} \mu}{2D_x} \left( \langle y^2\rangle - \frac{D_x}{\bar{k}} \right) + \frac{D_\kappa (1-\theta)}{D_x\mu} \langle y^2\rangle~.
        \end{aligned}\end{equation}
Again, we note that there is no direct contribution from the switching dynamics as $\varphi$ and $\psi$ are governed by equilibrium processes. We show in \cite{Note1} that
\begin{align}
    \langle y^{2}(t) \rangle 
    = 4 D_x \int_{-\infty}^t dt' \exp\left[ - 4\left(\bar{k} -  \frac{2D_\kappa(1+\theta)}{\mu^2}\right)(t-t')    \right. \nonumber \\
    \left.  + \frac{8D_\kappa (1+\theta)}{\mu^3} (e^{-\mu(t-t')}-1)  \right]~, \label{eq:fullvar}
\end{align}
which remains finite only if $2D_\kappa(1+\theta)< \bar{k}\mu^2$, in which case Eq.\,\eqref{eq:fullEP} can be computed exactly [see Fig.\,\ref{fig:AOUP_EP}(b)]. Consistently with the second law of thermodynamics, $\bar{k}\langle y^2\rangle \geq D_x$ \cite{Alston2022b}. 

We now consider the limit $\theta = -1$, which maximises how nonreciprocal the interaction fluctuations can be, generating the most interesting collective dynamics. Here, the total stiffness $\varphi(t) \rightarrow 0$ in a deterministic manner and the dynamics of both $y(t)$ and $\psi(t)$ reduce to independent, equilibrium diffusive processes in an external potential. The drift term for the center of mass $x(t)$ is the product of these two equilibrium processes \cite{Bothe2021,Bechinger2016}. 

The stationary probability distribution for the product $\omega=\psi y$ can be evaluated formally as $P_\omega \left( \omega \right) = \int_{-\infty}^\infty {d\omega'P_\psi(\omega/\omega') P_y(\omega')}/{|\omega'|}$ where $P_\psi$ and $P_y$ denote the Boltzmann steady-state probability densities of the corresponding (equilibrium) processes. We can then derive an expression for the stationary distribution for the drift $P_v(v=-\omega/2)$ through transformation of probability density functions. It reads
\begin{equation}\label{eq:Pv_V0}
	P_v \left(v = -\frac{\psi y}{2}\right) = \frac{1}{\pi} \sqrt{\frac{\bar{k}\mu}{D_x D_\kappa}} K_0\left[\sqrt{\frac{\bar{k}\mu}{D_x D_\kappa}} |v|\right]
\end{equation}
with $K_0$ the modified Bessel function of the second kind. 

In the present limit, the MSD for the center of mass $x(t)$ is easily calculated as the two-time correlators for $\psi(t)$ and $y(t)$ are those of an equilibrium OU process. Using Eq.\,(\ref{eq:MSD_general}), we obtain
\begin{align}\label{eq:fullmsd}
     \langle x^2(t)\rangle =&D_xt + \\
     &\frac{2D_\kappa D_x}{\mu\bar{k}(\mu+2\bar{k})^2}\bigg[e^{-(\mu+2\bar{k})|t|}-1 + (\mu+2\bar{k})t\bigg]   \nonumber~,
\end{align}
which exhibits the \textit{diffusive-ballistic-diffusive} scaling characteristic of active particles \cite{Bechinger2016}, as shown in Fig.~\ref{fig:AOUP_EP}(a). Comparing the form of this MSD to that of a general active particle \cite{Bechinger2016, Howse2007}, which we derive in  \cite{Note1}, we identify an effective self propulsion speed $v_0\equiv (D_\kappa D_x/\mu \bar k)^{1/2}$, persistence time $\tau_p \equiv (\mu+2\bar{k})^{-1}$ and bare diffusivity $D\equiv D_x/2$. At short timescales, $t \ll \tau_p$, the center of mass follows a diffusive motion with diffusion coefficient $D_x$. At long times, the dimer exhibits diffusive motion characterized by the long-time effective diffusion coefficient 
\begin{equation}\label{eq:msd_2p}
    D_{\rm eff}^{({\rm cont.})}= \lim_{t\rightarrow\infty} \frac{\langle x^2(t)\rangle}{2t} = \frac{D_x}{2}\left[1 + \frac{2D_\kappa}{\mu\bar{k}\left(\mu+2\bar{k}\right)}\right],
\end{equation}
which is strictly larger than the bare center of mass translational diffusivity when $D_\kappa > 0$, i.e. in the presence of fluctuations. For sufficiently strong fluctuations, specifically $2 D_\kappa > \mu\bar{k}(\mu+2\bar{k})$, this effective diffusivity can strikingly exceed that of a single particle. This is in stark contrast with the classical $1/N$ scaling for the diffusivity of $N$ identical Langevin processes interacting by equilibrium pair interactions and thus represents a genuinely nonequilibrium feature of the present model. Below, we show numerically that this result holds in the presence of repulsive interactions.

To quantify these nonequilibrium dynamics, we evaluate the rate of entropy production from Eq.\,(\ref{eq:fullEP}). For $\theta=-1$, we note that the variance of the interparticle displacement satisfies $\langle y^2\rangle = D_x/\bar{k}$, such that the only non-zero contribution to the entropy production rate comes from the center of mass dynamics $x(t)$. Using Eq.\,(\ref{eq:fullEP}), we write this as
\begin{equation}
  \left. \lim_{t\rightarrow\infty}\dot{S}_i^{({\rm cont.})}(t) \right|_{\theta = -1} = \frac{2D_\kappa}{\bar{k}\mu}~.
\end{equation}
We discuss the limit $\theta = 1$ in \cite{Note1}. In this case, the interactions are always reciprocal, but the fluctuations in the coupling strength alone are sufficient to drive the system out of equilibrium \cite{Alston2022b, Guyon2004,Zhang2017}.

\begin{figure}
    \centering
    \includegraphics[scale=1]{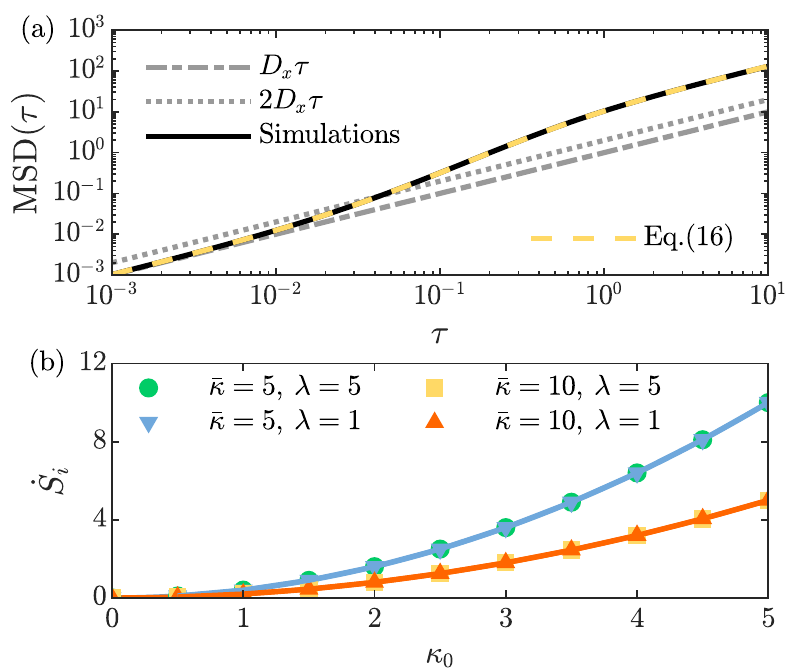}
    \caption{\textit{MSD and entropy production rate for synchronized discrete fluctuations ---}(a) MSD for the active bound states as given by Eq.\,(\ref{eq:fullMSD_RnT}) again displaying transient ballistic scaling and enhanced diffusion. Here $\kappa_0=5$ and $D_x=\bar k = \lambda = 1.$ (b) Rate of entropy production for the active bound state where we have fixed $D_x=1$. Symbols are measured from numerical simulations and solid lines are the result Eq.\,(\ref{eq:EPcom}).}
    \label{fig:Figure2}
\end{figure}

\textit{Discrete fluctuations in interaction potentials. ---} We now turn to the case of discrete stiffness fluctuations. We let $\kappa_{1,2}(t) \in \{-\kappa_0,+\kappa_0\}$ be correlated symmetric Telegraph processes \cite{Gardiner2009}; the symmetric nature of the Markov jump process ensures that $\langle \kappa_{1,2}(t)\rangle = 0$, and hence, that pair interactions remain on average reciprocal maintaining the particles in a bound state. The joint probability mass function ${\bf P}(t) = \big(P_{++}(t),P_{+-}(t),P_{--}(t),P_{-+}(t)\big)$ is generically governed by
\begin{equation}
\frac{d}{dt} {\bf P} = {\bf M} \cdot {\bf P}
\label{eq:meq_P}
\end{equation}
in which the position-independent Markov matrix ${\bf M}$ capturing the stochastic dynamics of the stiffnesses reads
\begin{equation}
    {\bf M}(\chi) = \lambda [(1-\chi) {\bf M}_{\rm mono} + \chi {\bf M}_{\rm bi}]
\end{equation}
with transition rate $\lambda > 0$, correlation parameter $\chi \in [0,1]$ and transition rate matrices defined as
\begin{equation}
    {\bf M}_{\rm mono} = 
    \begin{bmatrix}
    -2      & 1 & 0  & 1 \\
    1       & -2 & 1  & 0 \\
    0       & 1 & -2  & 1 \\
    1      & 0 & 1  & -2
\end{bmatrix},~
{\bf M}_{\rm bi} = 
    \begin{bmatrix}
    -1      & 0 & 1  & 0 \\
    0       & -1 & 0  & 1 \\
    1       & 0 & -1  & 0 \\
    0      & 1 & 0  & -1
\end{bmatrix}
\end{equation}

The limit $\chi = 0$ leads to bipartite dynamics, where each switching event causes a transition between reciprocal and nonreciprocal interaction. In contrast,  $\chi=1$ corresponds to maximally correlated switching dynamics, such that switching events are always synchronised and the (non)reciprocity of the dynamics is conserved by the fluctuations. While a study of the ensuing dynamics in the general case is of great interest, we consider here the limiting case $\chi=1$ only, such that ${\bf M} = \lambda {\bf M}_{\rm bi}$. In this limit, the fluctuations of $\varphi$ and $\psi$ are uncorrelated. 

Similarly to what was done for the continuous case, we further focus on the case where fluctuations are maximally nonreciprocal.
Let $\kappa_1(0) = - \kappa_2(0) \equiv \kappa_0$ at initialization, such that $|\psi(t)| = 2 \kappa_0$ and $\varphi(t)=0$. The synchrony condition $\chi = 1$ imposes that the total stiffness $\varphi(t)$ remains zero while the sign of the stiffness asymmetry $\psi(t)$ switches with symmetric Poisson rate $\lambda$, a Telegraph process, leading to nonreciprocal force fluctuations at all times (see details of allowed transitions for the Markov jump process in \cite{Note1}). The dynamics of $y(t)$ are exactly those of a diffusive particle in the potential $U (y)= \bar{k} y^2 $. We note that $\langle \psi(s)\psi(s')\rangle = 4\kappa_0^2 e^{-2\lambda|s-s'|}$, while $\langle y(s)y(s')\rangle=(D_x/\bar{k})e^{-2\bar{k}|s-s'|}$ is simply the propagator for an Ornstein-Uhlenbeck process \cite{Gardiner2009}. As shown in \cite{Note1}, we conclude that the full MSD then takes the form 
\begin{equation}\label{eq:fullMSD_RnT}
    \langle x^2(t)\rangle = D_x t + \frac{D_x\kappa_0^2}{2\bar{k}(\lambda+\bar{k})^2}\Big[e^{-2(\lambda+\bar{k})t}-1+ 2(\lambda+\bar{k})t\Big],
\end{equation}
[see Fig.~\ref{fig:Figure2}(a)] which again can be mapped to the MSD of an active particle with effective self-propulsion speed $v_0 \equiv \kappa_0 (D_x/\bar{k})^{1/2}$, persistence time $\tau_p \equiv (2(\lambda + \bar{k}))^{-1}$ and bare diffusivity $D \equiv D_x/2$ \cite{Howse2007, Bechinger2016, Note1}. Finally, the long-time effective diffusion coefficient reads
\begin{equation} 
    D_{\rm eff}^{({\rm disc.})} = \frac{D_x}{2} \left( 1 + \frac{\kappa_0^2}{\bar{k}\left(\lambda+\bar{k}\right)}  \right)~,
    \label{eq:rnt_longtime_D}
\end{equation}
which is strictly larger than the bare center of mass translational diffusivity. Remarkably, for sufficiently slow fluctuations, specifically $\lambda < \bar{k} (\kappa_0^2/\bar{k}^2 - 1)$, this effective diffusivity can exceed that of a single particle as we saw for the case of continuous fluctuations [see Fig.\,\ref{fig:Figure2}(a)].

As the dynamics for $y(t)$ is at equilibrium, the only non-zero contribution to the entropy production comes from the spontaneous drift of the center of mass. In the present case, the dynamics of $\psi(t)$ and $y(t)$ are again entirely decoupled implying that $\langle \psi^2 y^2\rangle =\langle \psi^2\rangle \langle y^2\rangle$. 
We can evaluate $\langle \psi^2\rangle = 4\kappa_0^2$, $\langle y^2\rangle = D_x/\bar k$ \cite{Gardiner2009} and using Eq.\,(\ref{eq:EP_general}) write the full entropy production rate as 
\begin{equation}\label{eq:EPcom}
    \lim_{t \to \infty} \dot{S}_i^{({\rm disc.})}(t) =\frac{2\kappa_0^2}{\bar k}.
\end{equation}
The independence of Eq.\,\eqref{eq:EPcom} on the switching rate $\lambda$ is demonstrated numerically in Fig.\,\ref{fig:Figure2}(b). The case of fluctuating reciprocal couplings, whereby we let $\kappa_1(0) = \kappa_2(0) = \kappa_0$ at initialization, is discussed in \cite{Note1}.

\textit{Effect of steric repulsion. ---} So far, we have ignored the role of steric repulsion; while this allowed us to derive exact analytical results, we now reintroduce a non-vanishing purely repulsive potential $U_r \neq 0$ in Eq.~\eqref{eq:2p_sys}. While the equation governing the center of mass dynamics is unaffected by this change, Eq.\,\eqref{eq:eom_y} for the interparticle displacement acquires an additional term
\begin{equation}
\dot{y}(t) = - (\varphi(t) + 2 \bar{k}) y(t) 
    - 2 \partial_y U_{r}(y) 
    + \sqrt{4 D_x} \xi_{y}(t)~.
\end{equation}
Intuitively, since $U_r$ should penalize particles overlapping, we expect the bound state to be characterized by a finite interparticle displacement, commensurate with the particle diameter. If we further assume that the fluctuations in $y$ are small, then $y(t) \approx \sqrt{\langle y^2\rangle}$ is approximately constant and $\psi(t)$ is left to be the sole term responsible for fluctuations in the self-propulsion contribution to the center of mass dynamics in Eq.~\eqref{eq:eom_x}. Remarkably, when $\psi(t)$ is an OU process, such as in Eq.\,(\ref{eq:varphi_dynamics_cont}), the ensuing dynamics of the bound state are then akin to those of an \textit{Active Ornstein-Uhlenbeck} (AOUP). Similarly, when $\psi(t)$ is governed by a Telegraph process, which is exactly the case which leads to Eq.\,(\ref{eq:fullMSD_RnT}), the dynamics match those of a  \textit{Run-and-Tumble} (RTP) particle.

\begin{figure}
    \centering
    \includegraphics[width=0.45\textwidth]{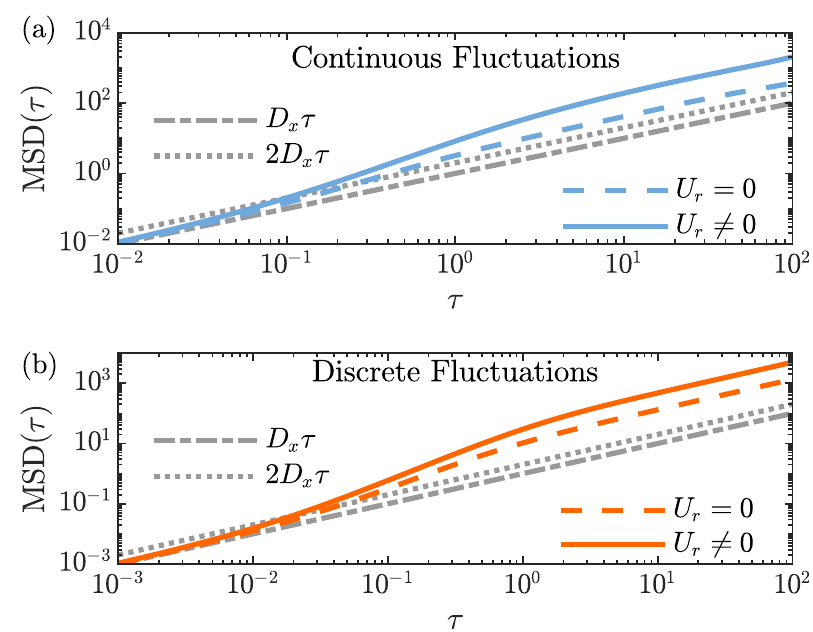}
    \caption{\textit{Steric repulsion enhances persistent motion ---} We compare our results on the dimer's dynamics above to simulations which include a repulsive force between the two particles through a WCA potential. For both (a) continuous and (b) discrete fluctuations, this additional interaction ensures that the ballistic scaling regime persists for longer and leads to a significantly enhanced effective diffusion coefficient. Parameters are the same as in Figs.\,1(a) and 2(a).}
    \label{fig:Figure3}
\end{figure}

To study numerically the effect that steric repulsion has on our results, we choose a Weeks-Chandler-Anderson potential for $U_{r}(r)$, capped at $r_c=2^{1/6}$ to give the particles a well-defined diameter \cite{Note1,Branka1999}. For the case of nonreciprocal fluctuations, the MSDs exhibit a longer phase of ballistic motion and an increased long-time diffusion coefficient, as can be seen in Fig.\,\ref{fig:Figure3}. The average rate of entropy production remains qualitatively unchanged by the introduction of steric repulsion showing the characteristic monotonic increase already seen in Fig.\,\ref{fig:Figure2} \cite{Note1}. By keeping the two particles apart, the size of the forces coming from the harmonic potential are on average much larger, driving more persistent motion and hence higher dissipation in the present model. 

\textit{Discussion and conclusion. --- }
We have demonstrated that reciprocal-symmetry-breaking fluctuations about a reciprocal mean attractive coupling are sufficient to generate two-particle bound states whose center of mass motion can be mapped onto that of a motile active particle, Eq.~\eqref{eq:eom_x}. For specific choices of the fluctuations and in the presence of steric repulsion, one-dimensional \textit{Active Ornstein-Uhlenbeck} as well as \textit{Run-and-Tumble} dynamics are recapitulated. We characterize the dissipative nature of these active bound states by computing the average rate of entropy production, Eq.~\eqref{eq:EP_general}. Remarkably, for sufficiently strong nonreciprocal fluctuations, the long time effective diffusion is observed to exceed that of a single particle, Eqs.~\eqref{eq:msd_2p} and \eqref{eq:rnt_longtime_D}, which represents a genuinely nonequilibrium feature of our model. 

Fluctuations in the degree of reciprocity of pair interactions arise naturally in a number of physical circumstances, e.g. through mediation by a nonequilibrium medium \cite{Durve2018,Hayashi2006,Saha2019,Soto2014,AgudoCanalejo2019}, in the presence of memory \cite{Loos2019} or from perception within a finite vision cone \cite{Loos2022,Lavergne2019}. In fact, in macroscopic active systems, nonreciprocity is arguably the norm rather than the exception. 

A simple model of fluctuating interactions in a two-body system was realized in a recent study on nano-particles subjected to random electric fields \cite{LuisHita2022}. The resulting pair-interactions were nonreciprocal in nature and lead to persistent motion as predicted in the theory above. Another promising candidate for realizing the dynamics studied here could be size-dependent interactions between droplets exchanging mass through inverse Ostwald ripening dynamics \cite{Theveneau2013}. In general, the strength of interactions through a medium depends on the perimeter of a droplet, which is a dynamic quantity due to thermal fluctuations driving constant re-balancing of the Laplace pressure. 

\begin{acknowledgments}
LC and HA contributed equally to this work. LC acknowledges support from the Francis Crick Institute, which receives its core funding from Cancer Research UK, the UK Medical Research Council and the Wellcome Trust (FC001317). HA was supported by a Roth PhD scholarship funded by the Department of Mathematics at Imperial College London.
\end{acknowledgments}


%
	
\end{document}


\title{Supplemental Material for ``Active Bound States Arise From Transiently Nonreciprocal Pair Interactions"}
	
	\author{Luca Cocconi}
	\thanks{These authors contributed equally.}
	\affiliation{Department of Mathematics, Imperial College London, South Kensington, London SW7 2AZ, United Kingdom}
	\affiliation{The Francis Crick Institute, London NW1 1AT, United Kingdom}
	\affiliation{Max Planck Institute for Dynamics and Self-Organization (MPIDS), 37077 G\"{o}ttingen, Germany}
		
	\author{Henry Alston}
	\thanks{These authors contributed equally.}
	\affiliation{Department of Mathematics, Imperial College London, South Kensington, London SW7 2AZ, United Kingdom}
	
	\author{Thibault Bertrand}%
	\email{t.bertrand@imperial.ac.uk}
	\affiliation{Department of Mathematics, Imperial College London, South Kensington, London SW7 2AZ, United Kingdom}
	
	\date{\today}

	\maketitle
 
 	\hrule
	\tableofcontents
 	\vspace{2em}
	\hrule	
	
\section{Change of frame of reference and entropy production rate}

\subsection{Change of frame of reference}
We consider a pair of Brownian particles with positions $x_1(t), x_2(t) \in \mathbb{R}$ and diffusivity $D_x$. Each particle is confined in a harmonic potential with time-dependent stiffness $k_i(t) = \bar k + \kappa_i(t)$ for $i=1, 2$ generated by the other particle. These dynamics thus combine a static, reciprocal attractive interaction between the particles (coming from the mean stiffness $\bar k > 0$) and a fluctuation contribution which may be non-reciprocal, that is $\kappa_1(t)$ and $\kappa_2(t)$ may be unequal at any given time. We argue that this constitutes the simplest model of interactions with reciprocal-symmetry-breaking fluctuations. \\
	
Realistic particles also commonly interact through a (reciprocal) short-range repulsive potential, $U_r(r)$, giving each of the particles a well-defined size. The dynamics of the two particle systems are not analytically tractable for general $U_r(r)$ and we therefore ignore this constraint for the time being.
	
In the overdamped limit, the governing equations then take the form
\begin{subequations}
	\begin{align}
		\dot{x}_1 &= -k_1(x_1 - x_2) + \sqrt{2 D_x}\xi_1 \\
		\dot{x}_2 &= -k_2(x_2 - x_1) + \sqrt{2 D_x}\xi_2,
	\end{align}
	\label{eq:2p_sys_SI}%
\end{subequations}
where $\xi_1(t)$ and $\xi_2(t)$ are independent, zero-mean, unit variance Gaussian white noise terms.

It is instructive to consider the dynamics of this two particle system in an alternative frame of reference by defining the center of mass $x=(x_1+x_2)/2$ and interparticle displacement $y=x_1-x_2$ coordinates. From Eq.\,(\ref{eq:2p_sys_SI}), we can then derive governing equations for the dynamics of these two variables:
 \begin{subequations}
 	\begin{align} 
 		\dot{x}(t) &= -\frac{1}{2} (\kappa_1-\kappa_2) y(t) + \sqrt{D_x/2}\big( \xi_1(t) + \xi_2(t)\big) \label{eq:eom_x_SI}\\
 		\dot{y}(t) &= - (\kappa_1+\kappa_2 + 2 \bar{k})y(t) + \sqrt{2 D_x}\big( \xi_{1}(t) -  \xi_{2}(t)\big) \label{eq:eom_y_SI}
 	\end{align}
 	\label{eq:2p_sys_ref_int}%
 \end{subequations}
The noise terms can be re-written succinctly each as a single Gaussian white noise term. Indeed, we define the noise terms  
\begin{equation}
	 \xi_x(t) = \frac{1}{2}\big(\xi_1(t) + \xi_2(t)\big)\quad\text{and}\quad \xi_y(t) = \frac{1}{2}\big(\xi_1(t) - \xi_2(t)\big),
\end{equation}
where we include the prefactors of $1/2$ such that both $\xi_x$ and $\xi_y$ have zero-mean and unit-variance, while also remaining independent:
\begin{equation}
\langle \xi_x(t)\rangle = \frac{1}{2}\big(\langle \xi_1(t)\rangle+\langle \xi_2(t)\rangle\big) = 0,\quad	\langle \xi_x(t)\xi_x(t')\rangle = \frac{1}{2}\big(\langle\xi_1(t)\xi_1(t')\rangle + \langle\xi_2(t)\xi_2(t')\rangle\big) = \delta(t-t')
\end{equation}
and similarly for $\xi_y(t)$:
\begin{equation}
	\langle \xi_y(t)\rangle =\frac{1}{2}\big(\langle \xi_1(t)\rangle-\langle \xi_2(t)\rangle\big) = 0,\quad	\langle \xi_y(t)\xi_y(t')\rangle = \frac{1}{2}\big(\langle\xi_1(t)\xi_1(t')\rangle + \langle\xi_2(t)\xi_2(t')\rangle\big) = \delta(t-t').
\end{equation}
Finally, we define the \textit{stiffness asymmetry} $\psi(t)=\kappa_1(t)-\kappa_2(t)$ and the \textit{total stiffness fluctuations} $\varphi(t)=\kappa_1(t) + \kappa_2(t)$, where we remark that $\psi(t)\ne 0$ is the signature of {\it broken reciprocal symmetry} at time $t$. 

In all, this leads us to re-writing Eq.\,(\ref{eq:2p_sys_SI}) in the new frame of reference as
\begin{subequations}
	\begin{align} 
		\dot{x}(t) &= -\frac{1}{2} \psi(t) y(t) + \sqrt{D_x} \xi_x(t) \label{eq:eom_x}\\
		\dot{y}(t) &= - (\varphi(t) + 2 \bar{k}) y(t) + \sqrt{4 D_x} \xi_{y}(t)~. \label{eq:eom_y}
	\end{align}
	\label{eq:2p_sys_ref_SI}%
\end{subequations}

\subsection{Full derivation of the steady-state entropy production rate}

Here we give in full the derivation of the steady-state entropy production rate for our two-particle system, which we write as the Kullback-Leibler divergence per unit time of the ensemble of forward $(x,y,\varphi,\psi)$ trajectories and their time-reversed counterparts,
\begin{equation}\label{eq:def_epr_kl}
    \lim_{t\to\infty} \dot{S}_i = \lim_{\tau \to \infty} \frac{1}{\tau} \left\langle \log \frac{\mathbb{P}_F[x,y,\varphi,\psi]}{\mathbb{P}_R[x,y,\varphi,\psi]} \right\rangle~,
\end{equation}
with $\mathbb{P}_F$ and $\mathbb{P}_R$ denoting corresponding path probabilities densities, while $\tau$ is the path duration. By straightforward manipulation of the joint path probabilities we now write
\begin{align}
    \mathbb{P}_{F,R}[x,y,\varphi,\psi] 
    &= \mathbb{P}_{F,R}[y,\varphi,\psi] \mathbb{P}_{F,R}[x|y,\varphi,\psi] \nonumber \\
    &= \mathbb{P}_{F,R}[y,\varphi] \mathbb{P}_{F,R}[\psi|y,\varphi] \mathbb{P}_{F,R}[x|y,\psi]
\end{align}
where in the second equality we have used that $x$ is independent of $\phi$ by Eq.~\eqref{eq:eom_x}. We further assume, as is the case in all models studied here, that $\psi$ is independent of $y$ and $\varphi$, whereby $\mathbb{P}_{F,R}[\psi|y,\varphi] = \mathbb{P}_{F,R}[\psi]$. Substituting back into Eq.~\eqref{eq:def_epr_kl} for the entropy production rate and writing logarithms of products as sums, we arrive by linearity of expectation at
\begin{equation} \label{eq:epr_decomp_kl}
    \lim_{t\to\infty} \dot{S}_i = 
     \lim_{\tau \to \infty} \frac{1}{\tau} \left\langle \log \frac{\mathbb{P}_F[\psi]}{\mathbb{P}_R[\psi]} \right\rangle + \lim_{\tau \to \infty} \frac{1}{\tau} \left\langle \log \frac{\mathbb{P}_F[y,\varphi]}{\mathbb{P}_R[y,\varphi]} \right\rangle + \lim_{\tau \to \infty} \frac{1}{\tau} \left\langle \log \frac{\mathbb{P}_F[x|y,\psi]}{\mathbb{P}_R[x|y,\psi]} \right\rangle~.
\end{equation}
The first term in the above vanishes when the dynamics of $\psi$ are of the equilibrium type and thus satisfy time-reversal symmetry, as is the case for all models studied here. The second term corresponds to the entropy production of marginal dynamics $(y,\varphi)$, which can be mapped onto diffusive motion in a stochastically evolving potential, as studied in \cite{Alston2022b}, from which expressions for this term (denoted $\dot{S}_i^{(y)}$ elsewhere for brevity) can be read off. Further insight into the last term in Eq.~\eqref{eq:epr_decomp_kl} is gained by expressing the conditional path probabilities for the $x$ dynamics as governed by the Langevin equation \eqref{eq:eom_x} in the Onsager-Machlup path integral formalism
\begin{subequations}\begin{align}
    \mathbb{P}_F[x|y,\psi] \propto {\rm exp}\left[ - \frac{1}{2D_x}\int_0^\tau dt \ \left( \dot{x} + \frac{y\psi}{2}\right)^{ 2} \right] \\
    \mathbb{P}_R[x|y,\psi] \propto {\rm exp}\left[ - \frac{1}{2D_x}\int_0^\tau dt \ \left( \dot{x} - \frac{y\psi}{2}\right)^{ 2} \right]
\end{align}\end{subequations}
where stochastic integrals are to be interpreted in the Stratonovich mid-point convention. Substituting into the last term in Eq.~\eqref{eq:epr_decomp_kl} we thus have
\begin{equation} \label{eq:epr_contrib_selfp}
    \lim_{\tau \to \infty} \frac{1}{\tau} \left\langle \log \frac{\mathbb{P}_F[x|y,\psi]}{\mathbb{P}_R[x|y,\psi]} \right\rangle = - \lim_{\tau \to \infty} \frac{1}{\tau D_x} \int_0^\tau dt \ \langle \dot{x}(t) \psi(t) y(t) \rangle = \frac{\langle \psi^2 y^2 \rangle}{2D_x}~.
\end{equation}
Eq.~\eqref{eq:epr_contrib_selfp} combined with Eq.~\eqref{eq:epr_decomp_kl} and the assumption that the $\psi$ dynamics are equilibrium, whereby as already mentioned the first term in Eq.~\eqref{eq:epr_decomp_kl} vanishes, amount to the expression for the entropy production used in the main text.

\section{Continuous stiffness fluctuations}

\subsection{Full derivation of time-dependent MSD}

Here we give the full derivation for the MSD for the case where the spatial dynamics are given by Eqs.\,(\ref{eq:2p_sys_ref_SI}) and the stiffness fluctuations $\kappa_i(t)$ are correlated Ornstein-Uhlenbeck processes with rate $\mu$ and diffusivity $D_\kappa$, 
\begin{equation}
	\dot{\kappa}_{i}(t) = -\mu \kappa_{i}(t)  + \sqrt{2D_\kappa}\bar{\eta}_{i}(t), \quad i \in \{1,2\}~, \label{eq:eom_AOU}
\end{equation}
with $\bar{\eta}_{1,2}(t)$ taken to be zero-mean white noises satisfying
\begin{equation}
	\langle \bar{\eta}_{i}(t)\bar{\eta}_{j}(t')\rangle = C_{ij} \delta(t-t'),\quad\mathrm{with} \quad {\bf C} = \begin{pmatrix} 1 & \theta \\ \theta & 1 \end{pmatrix}~.
\end{equation}
Here, ${\bf C}$ denotes the symmetric covariance and $\theta \in[-1, 1]$ quantifies how correlated the stiffness fluctuations are. The {stiffness asymmetry} $\psi(t)=\kappa_1(t)-\kappa_2(t)$ and the {total stiffness fluctuations} $\varphi(t)=\kappa_1(t) + \kappa_2(t)$ are then governed by
\begin{subequations}
\begin{align}
    \dot{\psi}(t) &= -\mu \psi(t) + \sqrt{4D_\kappa(1-\theta)} \eta_\psi(t) \\
    \dot{\varphi}(t) &= -\mu \varphi(t) + \sqrt{4D_\kappa(1+\theta)} \eta_\varphi(t) 
\end{align}
\end{subequations}
with $\langle \eta_i(t) \eta_j(t')\rangle = \delta_{ij} \delta(t-t')$.

Consider the dynamics for $y$ in the limit where $\theta = -1$. The governing equations take the form 
\begin{subequations}
\begin{align}
    \dot{x}(t) &= - \psi(t) y(t)/2 + \sqrt{D_x} \xi_x(t), \label{seq:eom_x}\\
    \dot{y}(t) &= -2\bar k y(t) + \sqrt{4D_x}\xi_y(t),\label{seq:eom_y}\\
    \dot{\psi}(t) &= - \mu \psi(t) + \sqrt{8D_\kappa} \eta_\psi(t)
\end{align}
\end{subequations}
with $\varphi(t) = 0$ provided the right initial conditions. Integrating in time, we derive the solution
\begin{equation}
    y(t) = y_0e^{-2\bar k t} + \sqrt{4D_x}\int_0^t ds\:\xi_y(s) e^{-2\bar k(t-s)}.
\end{equation}
Taking an average over the noise, we see that $\langle y(t)\rangle=y_0e^{-2\bar k t}$. To evaluate the mean squared displacement of the center of mass, $\langle (x(t)-x_0)^2\rangle$, we set $x_0=0$ and write 
\begin{equation*}\begin{aligned}
    \langle x^2(t)\rangle &= \bigg\langle \int_0^t ds \bigg(-\frac{1}{2}\psi(s)y(s)+\sqrt{D_x}\xi_x(s)\bigg)\int_0^t ds' \bigg(-\frac{1}{2}\psi(s')y(s') + \sqrt{D_x}\xi_x(s')\bigg)\bigg\rangle \\
    &= \int_0^t ds\int_0^tds' \frac{1}{4} \langle \psi(s)\psi(s')y(s)y(s')\rangle  + D_x\delta(s-s').
\end{aligned}\end{equation*}
As $y$ and $\psi$ are independent, we can write $\langle \psi(s)\psi(s')y(s)y(s')\rangle = \langle \psi(s)\psi(s')\rangle\langle y(s)y(s')\rangle$. We are thus left with computing the two correlators. 

Next, we derive an analytic expression for the time-correlation function $\langle y(t)y(t')\rangle$, working at steady-state. To do this, we set $t'>t$ and consider a second correlation function $C_y(t, t') = \big\langle (y(t)-\langle y(t)\rangle ) ( y(t') - \langle y(t')\rangle)\big\rangle$ which we can write as
\begin{equation}
\begin{aligned}
C_y(t, t') &= 4D_x\bigg\langle \int_0^t ds \xi_y(s)e^{-2\bar k(t-s)} \:\int_0^{t'}ds' \xi_y(s')e^{-2\bar k(t'-s')} \bigg\rangle \\
&= 4D_x\int_0^t ds \int_0^{t'} ds' e^{-2\bar k(t+t'-s-s')}\langle \xi_y(s)\xi_y(s')\rangle\\
 &= \frac{D_x}{\bar k} \big[e^{-2\bar k(t'-t)} - e^{-2\bar k(t+t')}\big].
\end{aligned}
\end{equation}
From the definition of $C_y(t, t')$, we have an analytic expression for the time correlation function $\langle y(t)y(t')\rangle$ as desired:
\begin{equation}
    \langle y(t)y(t')\rangle = \bigg[y_0^2 - \frac{D_x}{\bar k}\bigg] e^{-2\bar k(t+t')} + \frac{D_x}{\bar k}e^{-2\bar k |t-t'|}
    \label{eq:corr_y}
\end{equation}
where we have relaxed the condition that $t'>t$. Following the same procedure, we also derive an expression for $\langle \psi(t)\psi(t')\rangle$:
\begin{equation}
    \langle \psi(t)\psi(t')\rangle = \bigg[\psi_0^2 - \frac{4D_\kappa}{\mu}\bigg] e^{-\mu(t+t')} + \frac{4D_\kappa}{\mu}e^{-\mu |t-t'|}~.
\end{equation}

Given these two correlators, we can now obtain an exact expression for the MSD of the center of mass of the dimer
\begin{equation}\begin{aligned}
     \langle x^2(t)\rangle &= \int_0^t ds\int_0^t ds'\: \frac{D_\kappa D_x}{\mu\bar k}e^{-(\mu+2\bar k)|s-s'|} + D_x\delta(s-s')\\
     &=\frac{2D_\kappa D_x}{\mu\bar k(\mu+2\bar k)^2}\bigg[(\mu+2\bar k)t + e^{-(\mu+2\bar k)t}-1\bigg] + D_xt, \quad t>0.
\end{aligned}\end{equation}
Finally, we conclude on the effective diffusion coefficient
\begin{equation}
    D^{\rm (cont.)}_{\rm eff} = \lim_{t\rightarrow\infty} \frac{\langle x^2(t)\rangle}{2t} = \frac{D_x}{2}\bigg[1 + \frac{2D_\kappa}{\mu\bar k(\mu+2\bar k)}\bigg].
\end{equation}	

\subsection{Variance in particle position for an OU$^2$}

The position ${y}$ of an overdamped Brownian particle trapped in a harmonic potential whose stiffness fluctuates about its mean value $2\bar{k} > 0$ in the manner of a Ornstein-Uhlenbeck (OU) process is governed by the following stochastic process
\begin{subequations}\begin{align}
	\partial_t {y}(t) &= -(2\bar{k} + {\varphi}(t)) {y}(t) + \sqrt{4 D_x} \eta_{y}(t) \\
	\partial_t {\varphi}(t) &= - \mu \varphi(t) +\sqrt{4(1+\Theta)D_{\kappa}} \eta_{\varphi}(t)~,
\end{align}\end{subequations}
where the additive noises satisfy $\langle \eta_i(t) \eta_j(t') \rangle = \delta_{ij} \delta(t-t')$. We dub this stochastic process the {\it OU$^2$ process}.

The formal solution for ${y}$ is given by
\begin{equation}
	{y}(t) = \sqrt{4D_x} \int_{-\infty}^t dt' \ \eta_{y}(t') {\rm exp}\left[ - 2\bar{k}(t-t') - \int_{t'}^t dt'' \ \varphi(t'') \right]~.
\end{equation}
We square this equation and then average over noise realisations to obtain an expression for the dynamic variance:
\begin{align} \label{eq:2ndmom}
	\langle {y}^{2}(t) \rangle 
	&= 4D_x \int_{-\infty}^t dt' \ \int_{-\infty}^t d\tau' e^{- 2\bar{k}(t-t')- 2\bar{k}(t-\tau')} \left\langle \eta_{y}(t') \eta_{y}(\tau') {\rm exp}\left[ - \int_{t'}^t dt'' \ \varphi(t'') - \int_{\tau'}^t d\tau'' \ \varphi(\tau'') \right] \right\rangle \nonumber \\ 
	&=4D_x \int_{-\infty}^t dt' \ e^{- 4\bar{k}(t-t')} \left\langle {\rm exp}\left[ - 2\int_{t'}^t dt'' \ \varphi(t'') \right] \right\rangle
\end{align}
where we used 
\begin{equation}
	\left\langle \eta_{y}(t') \eta_{y}(\tau') {\rm exp}\left[ - \int_{t'}^t dt'' \ \varphi(t'') - \int_{\tau'}^t d\tau'' \ \varphi(\tau'') \right] \right\rangle = \delta(t'-\tau') \left\langle {\rm exp}\left[ - \int_{t'}^t dt'' \ \varphi(t'') - \int_{\tau'}^t d\tau'' \ \varphi(\tau'') \right] \right\rangle
\end{equation}
since $\eta$ and $\varphi$ are uncorrelated and thus the expectation factorises.
We now use a standard identity between the moment generating function of the random variable $- 2\int_{t'}^t dt'' \ \varphi(t'')$ and the exponential of the corresponding cumulant generating function, which in this case gives
\begin{equation} \label{eq:id_genfunc}
	\left\langle {\rm exp}\left[ - 2\int_{t'}^t dt'' \ \varphi(t'') \right] \right\rangle = {\rm exp}  \sum_{m=1}^\infty \frac{1}{m!} \left\langle \left( - 2\int_{t'}^t dt'' \ \varphi(t'') \right)^m  \right\rangle_c
\end{equation}
where the subscript $\langle \bullet \rangle_c$ denotes a cumulant/connected correlation. Since $\varphi(t)$ is a zero-mean OU process, at steady state all cumulants except the second vanish. The latter reads
\begin{equation}
	\langle \varphi(t_1) \varphi(t_2) \rangle_c = \frac{8(1+\Theta)D_{\kappa}}{\mu} e^{-\mu |t_1-t_2|}~.
\end{equation}
Thus,
\begin{align}
	\left\langle {\rm exp}\left[ - 2\int_{t'}^t dt'' \varphi(t'') \right] \right\rangle 
	&= {\rm exp} \left( \frac{8(1+\Theta)D_{\kappa}}{\mu} \int_{t'}^t dt'' \int_{t'}^t d\tau'' e^{-\mu |t''-\tau''|}  \right) \nonumber \\
	&= {\rm exp} \left( \frac{8(1+\Theta)D_{\kappa}}{\mu} \frac{1}{\mu^2} \left(\mu(t-t') -1 + e^{-\mu(t-t')} \right)  \right) \nonumber \\
	&= {\rm exp} \left( \frac{8(1+\Theta)D_{\kappa}}{\mu^2} (t-t') \right) {\rm exp} \left( \frac{8(1+\Theta)D_{\kappa}}{\mu^3} (-1 + e^{-\mu(t-t')}) \right)~.
\end{align}
Going back to Eq.~\eqref{eq:2ndmom}, our expression for the variance reduces to
\begin{align} \label{eq:int_eq_m2}
	\langle {y}^{2}(t) \rangle 
	&= 4D_x e^{-\frac{8(1+\Theta)D_{\kappa}}{\mu^3}} \int_{-\infty}^t dt' \ e^{- \left(4\bar{k} -  \frac{8(1+\Theta)D_{\kappa}}{\mu^2}\right)(t-t')   } {\rm exp} \left( \frac{8(1+\Theta)D_{\kappa}}{\mu^3} e^{-\mu(t-t')} \right)~,
\end{align}
which constitutes the key result of this section.
The integral in the expression above has a divergent contribution as $t'\to-\infty$ when the exponent of the first term in the integrand changes sign. In other words, a necessary and sufficient condition for the existence of the second moment ($\langle {y}^{2}(t) \rangle <\infty$) is 
\begin{equation}\label{eq:cond_exst_2}
\bar{k} > \frac{2(1+\Theta)D_\kappa}{\mu^2}~.
\end{equation}
The main result Eq.\,(\ref{eq:int_eq_m2}) can be integrated numerically to obtain the steady-state variance (in the limit $t\rightarrow \infty$) and hence the average rate of entropy production in the two-particle system. 

\subsection{Case of reciprocal fluctuations ($\theta=1$)}

For $\theta = 1$, $\psi(t) \rightarrow 0$ and at steady-state, the center of mass $x(t)$ is diffusive and its dynamics decouple from that of the interparticle displacement $y(t)$, which itself behaves as a Brownian particle in a fluctuating potential, $U_{\rm tot}(y, t) = (2\bar{k} + \varphi(t))y^2/2$.

{In other words,} $y(t)$ is subject to the action of a harmonic confining potential with a stiffness that itself follows an Ornstein-Uhlenbeck process with stiffness $\mu$ and mean $2\bar{k}$. This model was previously studied in \cite{Alston2022b}: the average rate of entropy production has only one non-zero contribution:
\begin{equation}\label{eq:ep_theta1}
    \lim_{t\rightarrow\infty}\dot{S}_i(t) =\lim_{t\rightarrow\infty}\dot{S}^{(y)}_i(t) = \frac{\mu\bar{k}}{2D_x}\bigg(\langle y^2\rangle - \frac{D_x}{\bar{k}}\bigg).
\end{equation}
This can be evaluated using the result of the previous section.

\section{Discrete stiffness fluctuations}

\subsection{Full derivation of time-dependent MSD}

\begin{figure}
    \centering
    \includegraphics[scale=0.5]{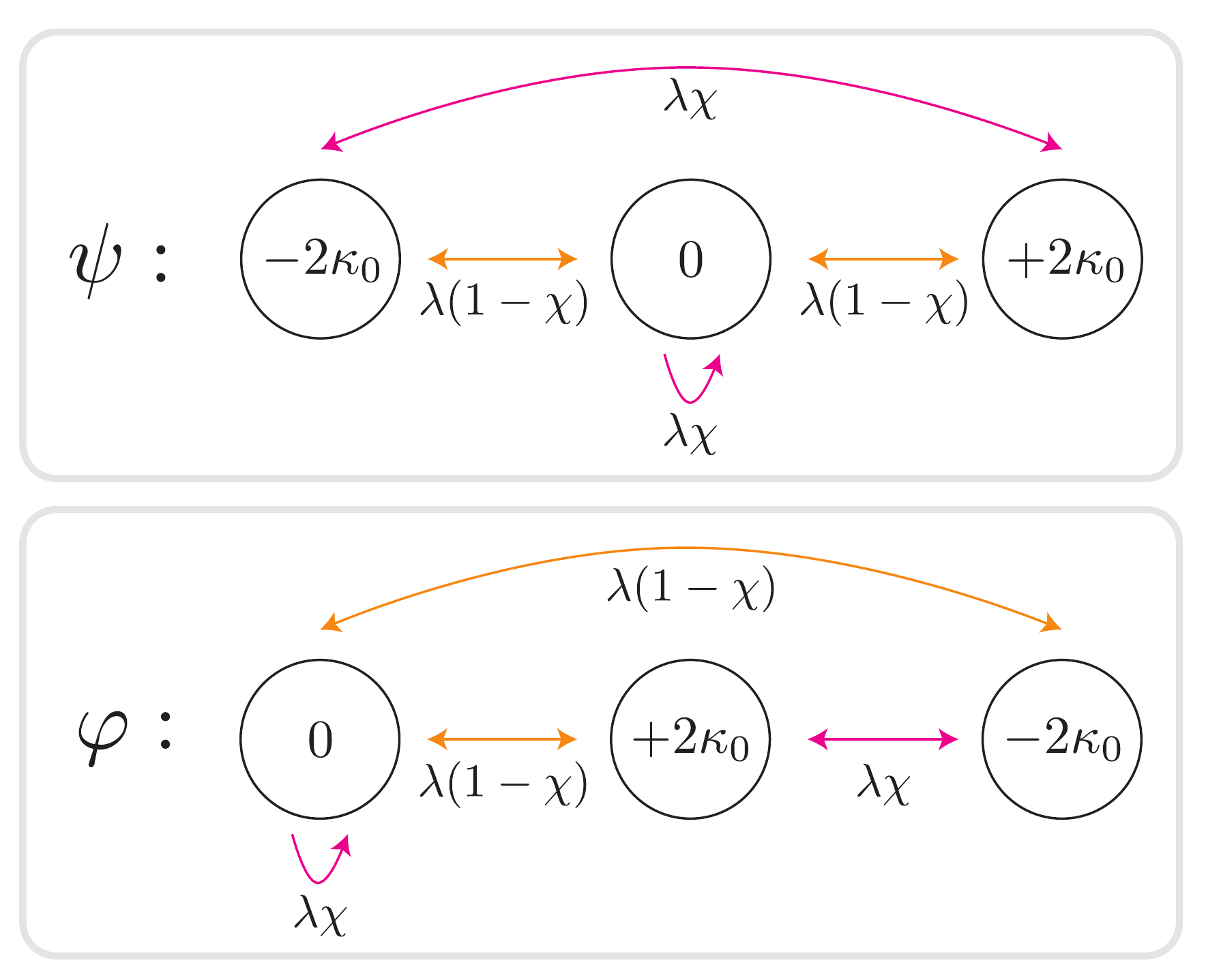}
    \caption{Allowed transitions for the Markov jump process controlling the stiffness asymmetry $\psi(t)$ and total stiffness fluctuation $\varphi(t)$ in the case of discrete fluctuations in the coupling strength. Here, $\lambda$ denotes the switching rate, while $\chi\in[0,1]$ is a correlation parameter controlling the degree of synchronisation of the single potential switching events.}
    \label{fig:schematic_NR_discrete}
\end{figure}

Here we give a full derivation of the time-dependent MSD calculated for the case of synchronised ($\chi=1$) discrete fluctuations in the main text.
The accessible transitions for the Markov processes governing the fluctuations of $\varphi$ and $\psi$ are shown in Fig.\,\eqref{fig:schematic_NR_discrete}. We know that the MSD takes the same general form in the 2-particle minimal model considered:
\begin{equation}\label{eq:MSD_general_SI}
    \langle x^2(t)\rangle = D_xt+ \frac{1}{4}\int_0^t ds\:\int_0^t ds' \: \langle \psi(s) \psi(s') \rangle \langle y(s) y(s') \rangle.
\end{equation}

Therefore, we are required to derive expressions for the two correlators. The stiffness asymmetry $\psi(t)$ is a telegraph process acting on $\{-2\kappa_0, 2\kappa_0\}$, thus we can write the correlator as \begin{equation}
    \langle \psi(s)\psi(s')\rangle = \langle\psi^2\rangle e^{-2\lambda|s'-s|} = 4\kappa_0^2 e^{-2\lambda|s'-s|}.
\end{equation}
where $\langle \psi^2\rangle = 4\kappa_0^2$ is the variance of $\psi$. In the absence of a repulsive potential, $U_{r}=0$, the dynamics for $y(t)$ are governed by an (equilibrium) Ornstein-Uhlenbeck process and using Eq.\,(\ref{eq:corr_y}), we write the MSD as
\begin{equation}
    \langle x^2(t)\rangle = D_xt+ \frac{D_x\kappa_0^2}{\bar k}\int_0^t ds\:\int_0^t ds' e^{-2(\lambda+\bar k)|s'-s|}.
\end{equation}
Finally, we use the fact that, for any constant $A_0>0$, 
\begin{equation}
    \int_0^t ds\:\int_0^t ds' e^{-A_0|s'-s|} = \frac{2}{A_0^2}\big[A_0t+e^{-A_0|t|}-1\big]
\end{equation}
to derive the final result 
\begin{equation}\begin{aligned}
    \langle x^2(t)\rangle= D_x t + \frac{D_x\kappa_0^2}{2\bar k(\lambda+\bar k)^2}\big[2(\lambda+\bar k)t 
    + e^{-2(\lambda+\bar k)|t|}-1\big]~,
\end{aligned}\end{equation}
which satisfies the usual \textit{diffusive-ballistic-diffusive} scaling observed with active particles. Indeed, at very short timescales, $t \ll (2(\lambda+\bar k))^{-1}$, the center of mass follows a diffusive motion with diffusion coefficient $D_x/2$. When $t\approx (2(\lambda+\bar k))^{-1}$, the two particle systems displays \textit{ballistic} motion, with $\langle x^2(t)\rangle= D_xt +(D_xk_0^2/\bar k)t^2$ and hence $\langle x^2(t)\rangle \propto t^2$. 
At large times, the effective diffusion coefficient reads
\begin{equation} \label{eq:rnt_longtime_D}
    D_{\rm eff}^{\rm (disc.)} = \frac{D_x}{2} \left( 1 + \frac{\kappa_0^2}{\bar k(\lambda+\bar k)}  \right)~,
\end{equation}
which is strictly larger than the bare center of mass translational diffusivity. Remarkably, for sufficiently slow fluctuations, specifically $\lambda < \bar{k} (\kappa_0^2/\bar{k}^2 - 1)$, this effective diffusivity can exceed that of a single particle. 

\subsection{Case of reciprocal fluctuations }
In the main text, we considered the scenario of synchronized discrete fluctuations with $\kappa_1(0) = \kappa_2(0) = \kappa_0$ at initialization, showing that this leads to persistent motion by deriving the time dependent MSD and the average rate of entropy production.

Now we consider again synchronized fluctuations in the coupling strength, but this time we suppose that $\kappa_1(0) = \kappa_2(0) = \kappa_0$ at initialization, leading to a vanishing stiffness asymmetry $\psi(t)=0$ while the total stiffness $\varphi(t) \in \{-2\kappa_0,2\kappa_0 \}$ is now governed by a Telegraph process with Poisson switching rate $\lambda$. The center of mass dynamics are here purely diffusive with $\dot{x}(t) = \sqrt{D_x} \xi_x$, leading trivially to a MSD $\langle x^2(t)\rangle = D_x t$, whereas the displacement dynamics are identical to those of a Brownian particle in a fluctuating harmonic potential $U_{\rm tot}(y,\varphi) = \big(2 \bar{k} + \varphi(t)\big) y^2/2 $. To ensure the existence of the second moment of the interparticle displacement, we require that $\kappa_0^2 < \bar{k}^2 + \lambda \bar{k}/2$~\cite{Guyon2004,Zhang2017}.

To obtain an explicit form for the entropy production in this case, we can use the result of \cite{Alston2022b} which states
\begin{equation}
    \lim_{t \to \infty} \dot{S}_i(t) = \frac{2\kappa_0^2 \lambda}{2\left(\bar{k}^2-\kappa_0^2\right)+\lambda\bar{k}}.
    \label{eq:EP_rec_dis}
\end{equation}
Hence, the second law of thermodynamics is satisfied if our earlier assumption $\kappa_0^2 < \bar{k}^2 + \lambda \bar{k}/2$ itself is.

\section{Effect of repulsive potential on rate of entropy production}
    In this section, we present numerical results for the entropy production rate in the two-particle system for the case where a repulsive potential is present between the two particles. We have defined the repulsive potential to be of Weeks-Chandler-Anderson form: 
    \begin{equation}
        U_{r}(r) = \frac{1}{12}\bigg[\bigg(\frac{\sigma}{r}\bigg)^{12} - \bigg(\frac{\sigma}{r}\bigg)^{6} \bigg],\quad r < 2^{1/6}\sigma
    \end{equation}
    and zero otherwise. We set $\sigma = 1$ in what follows, giving the particles an effective diameter of $r_c = 2^{1/6}$.  

\begin{figure}[h!]
        \centering
        \includegraphics[scale=0.9]{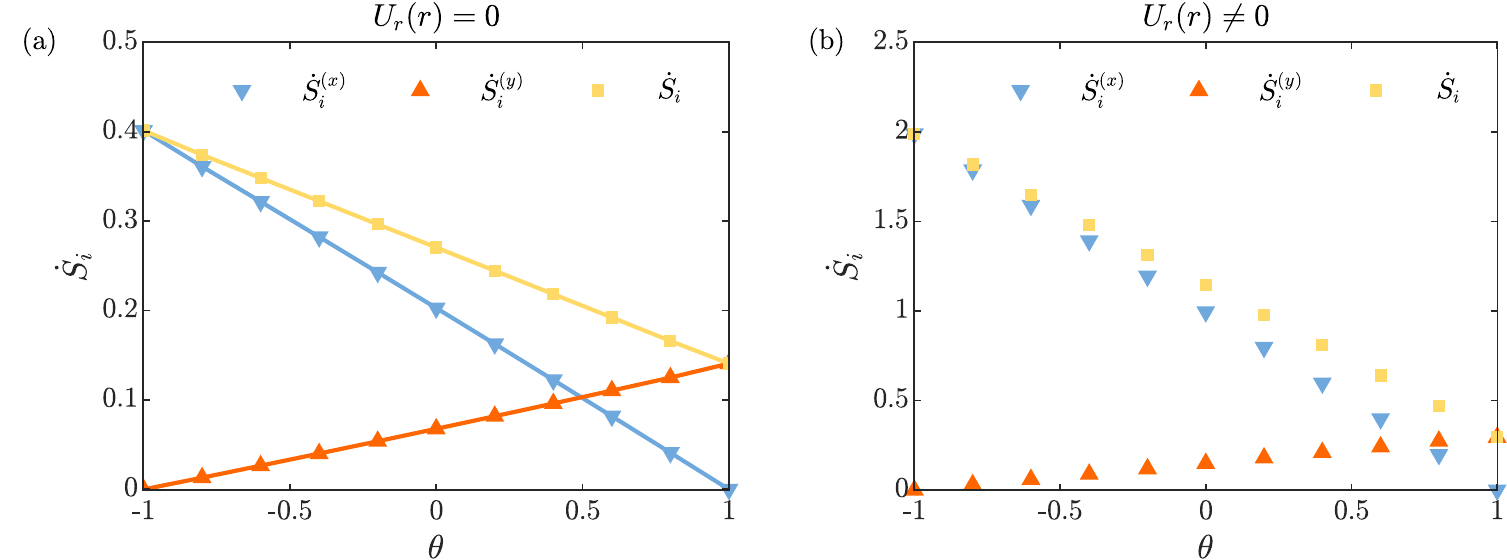}
        \caption{\textit{Entropy production rate when $U_{r}\ne 0$ for continuous fluctuations ---} (a) The entropy production rate when $U_{r}=0$ and (b) when the repulsive potential is of Weeks-Chandler-Anderson form. While the functional dependence on the correlation coefficient $\theta$ remains similar, the magnitude of all contributions is significantly larger. This agrees with the more persistent motion that we observe in Fig.\,3 in the main text. }
        \label{fig:FigureS2}
    \end{figure}

    In Fig.\,\ref{fig:FigureS2}, we see the effect for the continuous fluctuations. The linear scaling of the two contributions with the correlation coefficient $\theta$ persists, but the overall level of entropy production is approximately one order of magnitude higher. We argue that this is because the particle separation that is enforced by $U_{r}$ leads to greater forces stemming from the contribution of the harmonic potentials. These larger forces inevitably lead the potentials performing more nonequilibrium work, thus enhancing the entropy production. 
    
\begin{figure}[h!]
        \centering
        \includegraphics[scale=1.]{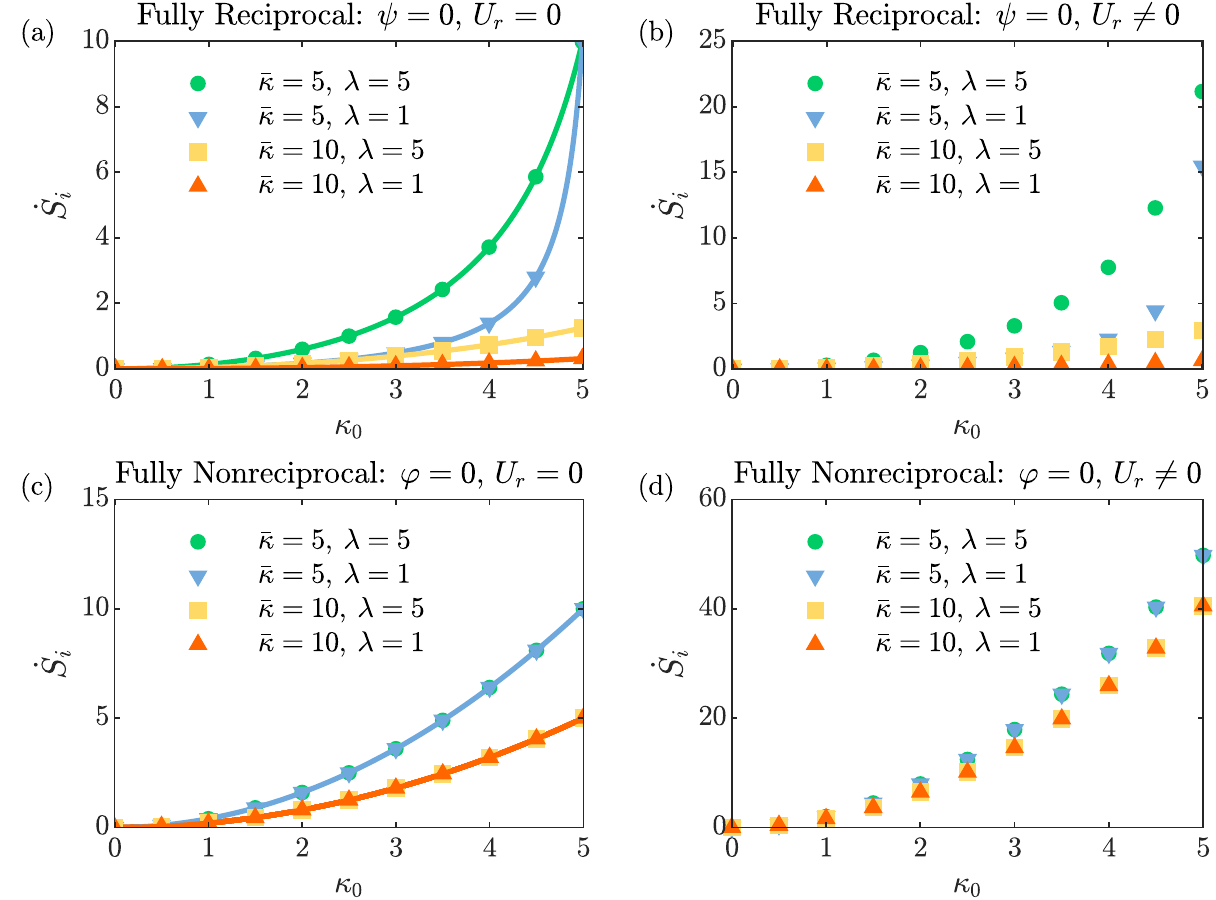}
        \caption{\textit{Entropy production rate when $U_{r}\ne 0$ for discrete fluctuations ---} (a) \& (b) compare the initialization which sets $\psi=0$ in the absence and presence of non-zero $U_{r}$, respectively. We observe a similar functional dependence on the fluctuation strength $\kappa_0$, but the overall amount of entropy produced is significantly increased, as was observed in the continuous case. (c) \& (d) The same is trend observed for the initialization which sets $\varphi=0$.  }
        \label{fig:FigureS3}
    \end{figure}

    In Fig.\,\ref{fig:FigureS3}, we study the case of discrete fluctuations and observe a similar trend. Notably, the entropy production rate is independent of the switching rate $\lambda$ for $\varphi=0$ as in the case $U_{r}=0$. This is due to the dissipation only depending on the square of the velocity so the sign switch, which is controlled by $\lambda$, doesn't factor in. The same is observed for a classic one-dimensional, symmetric \textit{Run-and-Tumble} particle, where the entropy production rate is independent of the Poissonian tumbling rate \cite{Seifert2012}.
    

\section{Comparison of results to MSDs of Generic Active Particle}
Here we re-derive a classic result for the mean-square displacement (MSD) of a generic self-propelling particle in 1D, then compare the structure of the full expression to the corresponding analytic forms derived for the two examples in the main text. 

Consider a self-propelled particle in 1D with the equation of motion
\begin{equation}
    \dot{x} = v(t) + \sqrt{2D_x}\eta(t)
\end{equation}
where $\eta(t)$ is a zero-mean, unit-variance Gaussian white noise term and $v(t)$ is a generic self-propulsion force. The MSD of the particle position can be then be derived as 
\begin{align}
    \langle x^2(t)\rangle &= \bigg\langle \bigg(\int_0^t ds \big[v(s)+\sqrt{2D}\eta(s)\big]\bigg) \bigg(\int_0^t ds' \big[v(s')+\sqrt{2D}\eta(s')\big]\bigg)\bigg\rangle \nonumber\\
    &= \int_0^t ds \int_0^t ds' \langle v(s)v(s')\rangle + 2D\delta(s-s').
\end{align}
To evaluate the leftover integrals, we require an analytic form for the time-correlation of the self-propulsion velocity. To make progress, we assume that the governing process for the self-propulsion force is time-translation invariant and that the decay of the correlation function is given or well approximate by an exponential, taking the form 
\begin{equation}
    \langle v(s)v(s')\rangle = v_0^2 e^{-|s-s'|/\tau_p}
\end{equation}
where we have defined the typical self-propulsion speed as 
\begin{equation}
    v_0 = \sqrt{\langle v^2(s)\rangle}
\end{equation}
(which is independent of s) and the persistence time $\tau_p$. We note that this form for the correlator is exact for both AOUP and RTP dynamics, where the governing processes for the self-propulsion force are Ornstein-Uhlenbeck and Telegraph processes, respectively. It would also capture the dynamics of an Active Brownian particle if we were working in higher dimensions (so the diffusion of the self-propulsion direction could be suitable defined). After substituting this expression for the correlator, we evaluate the integrals to derive the following result for the time-dependent MSD of a self-propelling particle 
\begin{equation}
    \text{MSD}(\tau) = 2D\tau + 2 v_0^2 \tau_p^2\bigg[\frac{\tau}{\tau_p} + e^{-\tau/\tau_p} - 1\bigg],
\end{equation}
with characteristic speed $v_0$, persistence time $\tau_p$ and bare diffusivity $D$ \cite{Howse2007, Bechinger2016}.

We use this general form to identify these three features of our active bound state dynamics. For the example in the main text with continuous stiffness fluctuations, we identify an effective self-propulsion speed $v_0 = (D_\kappa D_x/\mu \bar k)^{1/2}$, persistence time $\tau_p = (\mu+2\bar{k})^{-1}$ and bare diffusivity $D_x/2$. For the case of discrete fluctuations, we identify the characteristic self-propulsion speed $v_0 \equiv \kappa_0 (D_x/\bar{k})^{1/2}$, persistence time $\tau_p \equiv (2(\lambda + \bar{k}))^{-1}$ and bare diffusivity $D \equiv D_x/2$.

\section{Details of numerical analysis}
	We simulate the Langevin equations in the center of mass-interparticle displacement frame of reference using a stochastic Runga-Kutta solving method in 1D \cite{Branka1999}. We run the solver for different random number seeds for $10^3$ time units, only recording data after the first $20\%$ of the simulations as to let the dynamics reach a steady-state. 
	
	We measure the entropy production along each trajectory by calculating the heat dissipated at each step in the dynamics of the center of mass $x(t)$ and the interparticle displacement $y(t)$. This is given by the change in the variable multiplied by the effective (deterministic) force exerted on the variable at each timestep, evaluated in the Stratonovich convention for stochastic dynamics. For $x(t)$, the force is exactly the drift term identified in the main text: $v(t) = -\psi(t)y(t)/2$. Between $t$ and $t+dt$, we employ the Stratanovich convention for time discretization \cite{Seifert2012} which implies the heat dissipated $\delta Q$ is calculated as
	\begin{equation}
	    \delta Q([t, t+dt)) = \frac{v(t) + v(t+dt)}{2}[x(t+dt)-x(t)].
	\end{equation}
	We sum together all the contributions to measure the total heat dissipated across the trajectory. To recover the entropy production rate, we define the total change in entropy as $\delta S_i = \delta Q/T$ where $T$ is the effective temperature for the process. For the center of mass, we have $T=D_x/2$ which comes form the governing Langevin equation for $z(t)$. Finally, the \textit{rate} of entropy production is the change in entropy over the total length (in time) of the trajectory. We use the same method to calculate the entropy production from the interparticle displacement dynamics and show good agreement in all cases with our analytic results.
	
	We measure the MSD for the center of mass $\langle (x(t)-x(0))^2\rangle$ by recording $x(t)$ every $10^{-3}$ time units. We then let $dt$ be a multiple of $10^{-3}$ that is less than the total simulation time $T_{\rm traj}$ and evaluate the average value of $(x(t+dt)-x(t))^2$ for all $t \in [0, T_{\rm traj}-dt]$ such that $t$ is also a multiple of $10^{-3}$. The effective diffusion coefficient is then given by 
	\begin{equation}
	    D_{\rm eff} = \lim_{t\rightarrow\infty}\frac{\langle (x(t)-x(0))^2\rangle}{2t}.
	\end{equation}
	
	
%